# Interpretable machine learning-guided design of Fe-based soft magnetic alloys

Aditi Nachnani,[1] Kai K. Li-Caldwell,[1] Saptarshi Biswas,[1] Prince Sharma,[1]

Gaoyuan Ouyang,[1] and Prashant Singh[1,*]

[1]Ames National Laboratory, U.S. Department of Energy, Iowa State University, Ames, IA 50011, USA

**Abstract**

We present a machine-learning guided approach to predict saturation magnetization ($M_S$) and coercivity ($H_C$) in Fe-rich soft magnetic alloys, particularly Fe-Si-B systems. ML models trained on experimental data reveals that increasing Si and B content reduces $M_S$ from 1.81T (DFT~2.04 T) to ~1.54 T (DFT~1.56T) in Fe-Si-B, which is attributed to decreased magnetic density and structural modifications. Experimental validation of ML predicted magnetic saturation on Fe-1Si-1B (2.09T), Fe-5Si-5B (2.01T) and Fe-10Si-10B (1.54T) alloy compositions further support our findings. These trends are consistent with density functional theory (DFT) predictions, which link increased electronic disorder and band broadening to lower $M_S$ values. Experimental validation on selected alloys confirms the predictive accuracy of the ML model, with good agreement across compositions. Beyond predictive accuracy, detailed uncertainty quantification and model interpretability including through feature importance and partial dependence analysis reveals that $M_S$ is governed by a nonlinear interplay between Fe content, early transition metal ratios, and annealing temperature, while $H_C$ is more sensitive to processing conditions such as ribbon thickness and thermal treatment windows. The ML framework was further applied to Fe-Si-B/Cr/Cu/Zr/Nb alloys in a pseudo-quaternary compositional space, which shows comparable magnetic properties to NANOMET ($Fe_{84.8}Si_{0.5}B_{9.4}Cu_{0.8}P_{3.5}C_1$), FINEMET ($Fe_{73.5}Si_{13.5}B_9Cu_1Nb_3$), NANOPERM ($Fe_{88}Zr_7B_4Cu_1$), and HITPERM ($Fe_{44}Co_{44}Zr_7B_4Cu_1$. Our fundings demonstrate the potential of ML framework for accelerated search of high-performance, Co- and Ni-free, soft magnetic materials.

**Keywords**: Soft-magnets, Fe-alloys, AI/ML, Interpretable ML, Uncertainty quantification, DFT, VSM

*Corresponding author Email: psingh84@ameslab.gov/prashant40179@gmail.com

1. **Introduction:**

The quest for greater efficiency in energy conversion and transformation has driven the development of advanced energy materials, with soft magnetic materials playing a pivotal role [**1-5**]. These materials, characterized by their low coercivity and ability to rapidly respond to magnetic fields, are essential for minimizing energy loss and directing magnetic flux in electromagnetic devices. Their applications range from transformers and inductors to motors and generators, making them indispensable components in modern energy and transportation technologies [**6-9**].

Iron-based soft magnetic materials, including silicon steels, non-oriented steels, amorphous alloys, and nanocrystalline materials, exhibit tailored properties for diverse applications. Silicon steels, widely used in transformer cores, achieve saturation magnetizations of 2.0-2.1 T and coercivity values of 100-1000 A/cm [**3**], with performance closely tied to their grain-oriented structure that minimizes magnetic anisotropy along specific crystallographic directions. Non-oriented steels, designed for isotropic magnetic performance in rotating machinery, feature grain sizes of 10-50 μm, with coercivity values of 500-1500 A/cm and comparable saturation magnetization [**10**]. Amorphous materials, such as Fe-based metallic glasses, produced via rapid solidification, offer saturation magnetizations of 1.2-1.6 T and coercivity below 100 A/cm due to their disordered atomic structure [**11**], making them ideal for high-frequency applications. Nanocrystalline alloys combine ultra-fine grain sizes of 10-20 nm with saturation magnetizations of 1.2-1.3 T and coercivity values below 100 A/cm [**12**], benefiting from engineered grain boundaries and minimized anisotropy for high-frequency and thermal stability.

On the other hand, amorphous materials, such as Fe-based metallic glasses, already excel in high-frequency applications with low coercivity and high saturation magnetization, but developing ways to further enhance their thermal stability and resistance to mechanical stress will be critical for more demanding environments [**13**]. Fe-based alloys exhibit impressive soft magnetic properties with ultra-fine grain sizes, low coercivity, and high magnetic saturation [**14-16**], but future material designs must aim to push these properties further, addressing the need for enhanced stability. The future design of soft magnetic materials hinges significantly on the development of new compositions that can optimize key properties such as saturation magnetization, coercivity, permeability, and core loss. One avenue of improvement lies in refining the alloy compositions used in current materials, such as silicon steels, non-oriented steels, and nanocrystalline alloys [**17**].

In this work, we build two machine-learning models for predicting saturation magnetization ($M_s$) and coercivity ($H_c$) and relate these magnetic properties to compositions of new soft-magnetic materials. Both the models were trained on experiment database acquired from the literature comprising of Fe-rich soft-magnetic materials. In our assessment, we primarily focused on Fe-Si-B-based soft-magnetic alloys for their uses in electrical and magnetic applications due to their excellent magnetic properties, including high permeability, low coercivity, and low core loss. The silicon is known to improve the electrical resistivity while boron enhances the amorphous structure of the material. Notably, we avoided Ni and Co in the search of new soft-magnetic materials due to their criticality from a supply-chain perspective, making their minimization beneficial for material sustainability and cost efficiency. Elements such as niobium (Nb), and zirconium (Zr) are incorporated to refine grain boundary characteristics and minimize anisotropy [18], leading to higher saturation magnetization and lower coercivity [18]. Additionally, metalloids like silicon (Si) and boron (B) are introduced to promote glass formation [19], while noble metals such as copper (Cu) act as nucleating agents for the nanocrystalline ferromagnetic phase [20,21], enabling the development of high-performance soft magnetic materials [22]. These alloys represent the cutting edge of soft magnetic technology for modern electromagnetic applications.

To further understand the fundamental electronic mechanisms governing soft-magnetic properties in Fe-Si-B/Cu/Nb/Zr, we employed density functional theory (DFT) to analyze electronic-structure, phase-stability and mechanical properties of selected set of alloying compositions. Our study provides some key electronic-structure insights into the features controlling $M_S$ and $H_C$, supporting the machine-learning predictions. The results demonstrate that the ML model, trained on experimental datasets and atomic/alloy features, successfully captures $M_S$ and $H_C$ trends, and presents a high-throughput framework for designing next-generation soft magnetic materials, accelerating the discovery of novel alloys for technological applications.

## 2. Methods

**Figure 1** shows the proposed ML framework for soft-magnetic material design, which are becoming an important and powerful tool for material discovery and design, even building emerging materials intelligence ecosystems [23-27]. The workflow consisting of four key stages: data acquisition and preprocessing, feature extraction and analysis, ML modeling, and predictions and insights. Notably, the predictions and insights stage leverage trained models to predict magnetic properties, generate contour plots, and establish correlations between experimental and predicted values, enabling data-driven optimization of material properties.

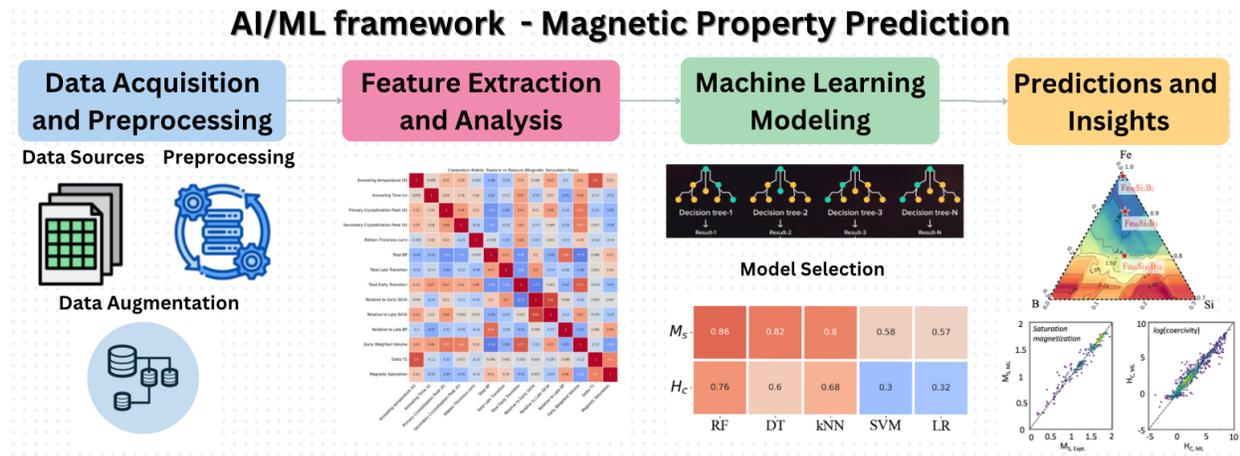

**Figure 1.** AI/ML framework for soft-magnetic material design.

**Data acquisition and preprocessing:** This work mainly focuses on $M_S$ and $H_C$ while searching for new soft-magnetic materials. Saturation magnetization and coercivity are critical properties in soft magnetic materials because they directly influence their performance in electromagnetic applications. High $M_S$ allows the material to achieve greater magnetic flux density while low $H_C$ is equally important because it determines how easily the material can be magnetized and demagnetized. We utilized $M_S$ and $H_C$ database provided from the Arroyave Lab [**17**] and curated some (mainly Fe-Si alloys) from various literature publications [**28-109**], and specifically to create machine-readable database of Fe-Si based soft-magnetic materials.

**Feature Engineering**: We approached our model development including extended feature list to achieve higher accuracy without bias and overdependence on specific features. The mean, variance, minimum and maximum values of each of those atomic, electronic and material features was added, and model was trained using only these feature sets. The mean and variance were both weighted according to the proportion of the element within each alloy structures. Mathematically, the mean ($\mu$), variance ($\sigma^2$), minimum (min), and maximum (max) values of a feature $X = \{x_1, x_2,\ldots,x_n\}$ are calculated as, $\mu = \frac{1}{n}\sum_{i=1}^{n} x_i$; $\sigma^2 = \frac{1}{n}\sum_{i=1}^{n}(x_i - \mu)^2$; $\min(X) = \min\{x_1, x_2, x_3, \ldots, x_n\}$; and $\max(X) = \max\{x_1, x_2, x_3, \ldots, x_n\}$.

To streamline the model and simplify its structure without losing critical information, we implemented a systematic five-step feature selection process. This involved removing features with over 50% missing values, eliminating features with only one unique value, identifying and discarding collinear features based on the Pearson correlation coefficient, and using a gradient boosting decision tree algorithm to remove features with zero importance and cumulative low importance [**110**]. The gradient boosting ML scheme combines multiple weak learners, i.e., decision trees, into a strong predictor by constructing trees sequentially, with each tree correcting the errors of the previous one [**111**]. Features contributing more to the decision splits at higher levels of the tree were considered more important, and their relative importance was averaged over a sequence of trees to reduce estimation variance. Features with zero importance were removed first, followed by those with low cumulative importance that were unnecessary for reaching a user-defined total feature importance threshold, such as 99%. After feature selection, the final MS model used 10 elemental features, while the HC model retained 59 features that included elemental inputs along with heat-treatment parameters (e.g., annealing temperature, crystallization peaks) and microstructural descriptors (e.g., weighted volume/mass, ribbon thickness, and crystallization-related metrics). This approach optimized the balance between model simplicity and predictive accuracy by retaining only the most relevant features [**112**].

**Model evaluation:** Evaluating the performance of machine learning models is a critical step in assessing their effectiveness. Here we have considered 4 parameters- R-squared ($R^2$) score, Mean Absolute Error (MAE), Root Mean Squared Error (RMSE), and Mean Absolute Percentage Error (MAPE), for evaluating our models' performance (**Eq. 1-4**). The R-squared score, also known as the coefficient of determination, first introduced by Wright [**113**] measures the proportion of the variance in the target variable that is explained by the model i.e., it shows how well the dependent variable is evaluated by all the independent variables [**114**]. It varies from 0 to 1 where a higher $R^2$ score means a better fit of the dataset. MAE is straightforward and measures the average absolute deviation between the model's predicted values and the actual target values. RMSE penalizes heavily to outliers thus making it more sensitive to outliers in a dataset, whereas MAPE focuses on percentage error and becomes effective in quantifying the relative variations of the predicted data and actual data, but it is ineffective when large errors are determined [**114**].

$$R2 = 1 - \frac{\sum (y_i - \hat{y})^2}{\sum (y_i - \overline{y})^2} \tag{1}$$

$$MAE = \frac{1}{N} \sum_{i=1}^{N} |y_i - \hat{y}| \tag{2}$$

$$RMSE = \sqrt{\frac{1}{N} \sum_{i=1}^{N} (y_i - \hat{y})^2} \tag{3}$$

$$MAPE = \frac{1}{N} \sum_{i=1}^{N} \frac{|y_i - \hat{y}|}{|y_i|} \times 100\% \qquad (4)$$

where $\bar{y}$ is the mean value of y, $\hat{y}$ is the predicted value of y, and $y_i$ is the $i^{th}$ data point.

**Experimental methods:**

*Materials:* The Fe and Si (99.9% purity) were procured from the Materials Preparation Center (Ames Lab), while B (99.5% purity) was obtained from Alfa Aesar. High-purity Ar (Matheson) was used for arc melting.

*Alloy Preparation:* Three compositions of FeSiB were selected, $Fe_{98}Si_1B_1$, $Fe_{90}Si_5B_5$, and $Fe_{80}Si_{10}B_{10}$. Approximately 10 grams of each composition were prepared using an arc melter. The total initial mass of the samples was recorded before melting. The arc melter was flushed with high purity argon three times and then kept at 2/3 atm. While melting, the power supply was set to 70 Amps. When preparing the ingots, the samples were melted, cooled, flipped, and then melted again a total of 3 times to ensure chemical homogeneity within the samples. After arc melting, initial mass of elements was compared to the final ingot mass to ensure that mass loss was under 0.5%. Ingots were then cut using a wire EDM(SYJ-7720) into needles with approximate dimensions of 1mm x 1mm x 4mm.

*Measurement of Saturation Magnetization:* The masses of each needle were measured and then prepared for VSM (DynaCool) measurement of saturation magnetization of each sample. The H field magnitude was set to 3 Tesla. Inside the chamber the pressure was kept at a low vacuum and run at 300K. The H and B fields were recorded as the applied field was scanned from -3T to +3T.

**Magnetic saturation calculation:** The saturation magnetization is determined from theoretical magnetization data given in Bohr magnetons per atom as

$$M_{S,DFT} = \frac{N_A \, \bar{\mu}}{V_m};$$

where $\bar{\mu}$, $N_A$, and $V_m$ in Eq. (1) are average magnetic moments of the alloy, Avogadro's number, and the molar volume, respectively, of Fe-Si-B alloy compositions [**115**]. See details of DFT method in supplemental file.

### 3. Results and Discussion:

The initial step involved performing a correlation analysis to select features, aiming to minimize computation time and enhance the model's robustness by eliminating redundant features [**17,116**]. To identify key features influencing magnetic saturation ($M_S$) and coercivity ($H_C$), Pearson correlation analysis was performed across 36-37 physical and compositional parameters (see **Figure S2**). For MS, strong positive correlations were observed with annealing temperature (0.61) and annealing time (0.58), highlighting the role of thermal processing in promoting grain growth and reducing structural disorder. Early-stage crystallization indicators such as Early Weighted Volume (0.54), Area (0.56), and Mass (0.55) also positively correlate with MS, suggesting that early nucleation supports uniform magnetic phase distribution. In contrast, MS is negatively correlated with features linked to late-stage crystallization, including Late Weighted Volume (-0.45) and Delta T2 (-0.50), implying that prolonged crystallization may introduce non-magnetic phases or structural inhomogeneities detrimental to saturation.

Coercivity shows an inverse dependence on several of the same features, see **Figure S3**. It correlates negatively with annealing temperature (-0.57) and annealing time (-0.49), indicating that longer thermal treatments relieve internal stress and reduce magnetic hardness. However, HC positively correlates with Delta T1 (0.44), Delta T2 (0.47), and Late Weighted Mass (0.50), suggesting that late-stage crystallization and larger thermal gradients increase magnetic hardness, likely due to phase coarsening or domain wall pinning. Comparative analysis reveals that early-stage crystallization enhances MS and suppresses HC, whereas late-stage crystallization has the opposite effect, underscoring the importance of finely tuned thermal control in optimizing soft magnetic performance.

**Model training, testing, and cross-validation details**: The heat-map in **Fig. 2** illustrates the performance of various ML models in predicting $M_S$ and $H_C$, evaluated based on the $R^2$ that measures how well the model's predictions align with the actual data, with values closer to 1 indicating better predictive accuracy. **Figure 2a** shows a comparative evaluation of different machine learning (ML) models—Random Forest (RF), Decision Tree (DT), k-Nearest Neighbors (kNN), Support Vector Machine (SVM), and Linear Regression (LR)—based on their predictive performance for $M_S$ and HC. The random forest models were implemented using scikit-learn with 1000–2000 estimators ($n$ estimators), max features set to 'sqrt', and min sample leaf of 1 to ensure model flexibility and sensitivity to subtle data variations. These hyperparameters were selected through empirical tuning to achieve optimal performance on the training data. The color-coded heatmap in **Fig. 2** represents the coefficient of determination ($R^2$), which quantifies how well the model captures the experimental trends. RF achieves the highest accuracy for both $M_S$ ($R^2$= 0.86) and $H_C$ ($R^2$= 0.76), making it the most reliable model for predicting magnetic properties. In contrast, SVM and LR perform poorly, especially for coercivity, with $R^2$ values of 0.3 and 0.32, respectively, suggesting their limitations in capturing the nonlinear relationships inherent in the dataset.

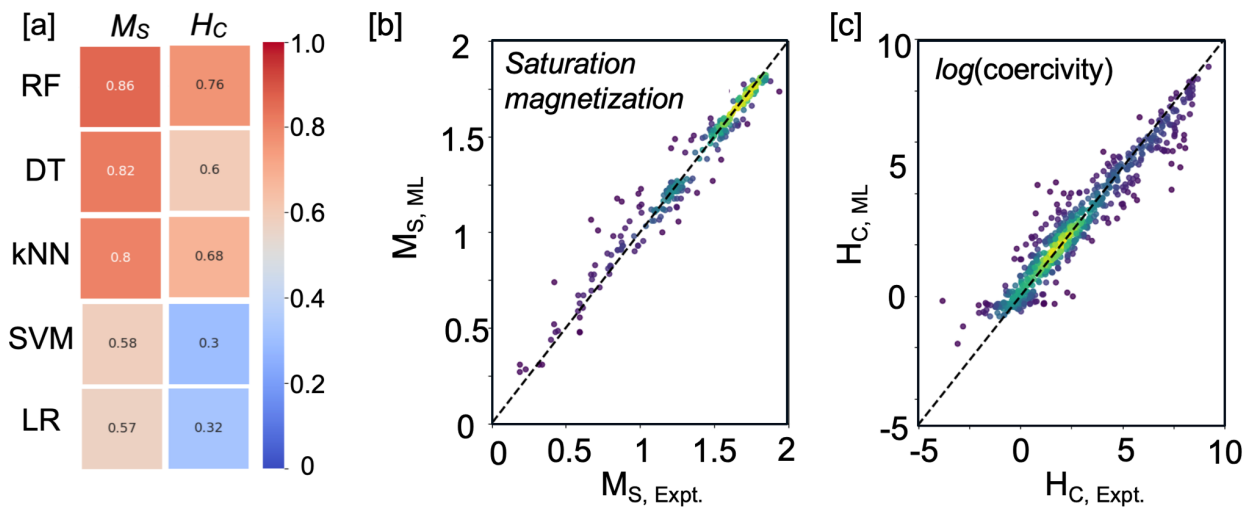

**Figure 2.** (a) The coefficient of determination ($R^2$) value of five different machine learning algorithms trained for magnetic saturation ($M_S$) and coercivity ($H_C$). The models considered in this study includes random forest (RF), decision tree (DT), k-nearest neighbors (kNN), support vector machines (SVM), and linear regression (LR). The parity plot for predicted vs experimental (b) $M_S$, and (c) $H_C$ of the best performing random-forest ML model from (a), showing strong agreement with experiments.

In **Fig. 2b&2c**, we show parity plots of our best predictive models, i.e., random forest ML models. The high $R^2$ for RF models show strong predictive performance both for saturation magnetization ($M_S$) and logarithmic coercivity ($log(H_C)$) The close alignment of predicted and experimental values in **Fig. 2b** confirms the model's accuracy for $M_S$, while **Fig. 2c** shows that the logarithmic transformation of $H_C$ improves prediction stability. Despite some scatter at higher $H_C$ values, RF effectively captures key trends, making it a best model for the high-throughput magnetic property predictions in Fe-rich soft-magnetic materials. All the models use similar set of features described in **Fig. S1&S2**. For the MS model, additional sample weights were applied to a small subset of data points representing critical compositions (e.g., low-Si/B alloys) to enhance sensitivity to trends in those regions. This worthwhile to mention that a weight of 2.0 was assigned to the lowest Si/B compositions (while the rest were 1.0) in the $M_S$ model (see SI **Fig. S4**) to enhance learning in underrepresented regions.

**Feature importance and uncertainty quantification of RF model (training set) for saturation magnetization**: In the section, we performed feature importance analysis and uncertainty quantification of the RF model in predicting magnetic saturation. **Figure 3a** ranked feature importance values derived from a Random Forest regression model trained to predict magnetic saturation in Fe-based alloys. The most influential feature is "Relative to Fe Early," which accounts for approximately 37.8% of the total predictive contribution, indicating that early transition elements relative to Fe have a strong effect on magnetic saturation. This is followed by "Annealing temperature (K)" at around 26.7%, suggesting that thermal processing conditions significantly impact the magnetic properties. "Fe" content itself contributes approximately 11.2%, confirming that composition plays a direct role but is not the sole determinant. Remaining features such as "Delta T1," "Total Early Transition," and "Coercivity" have lesser but non-negligible influence, each contributing below 4% to the model's decision process. This distribution emphasizes the combined effect of composition and thermal history on the magnetic behavior of the materials. While residual plot in **Fig. 3b** shows the difference between the actual and predicted magnetic saturation values as a function of the predicted values, which shows the accuracy of RF model without significant bias.

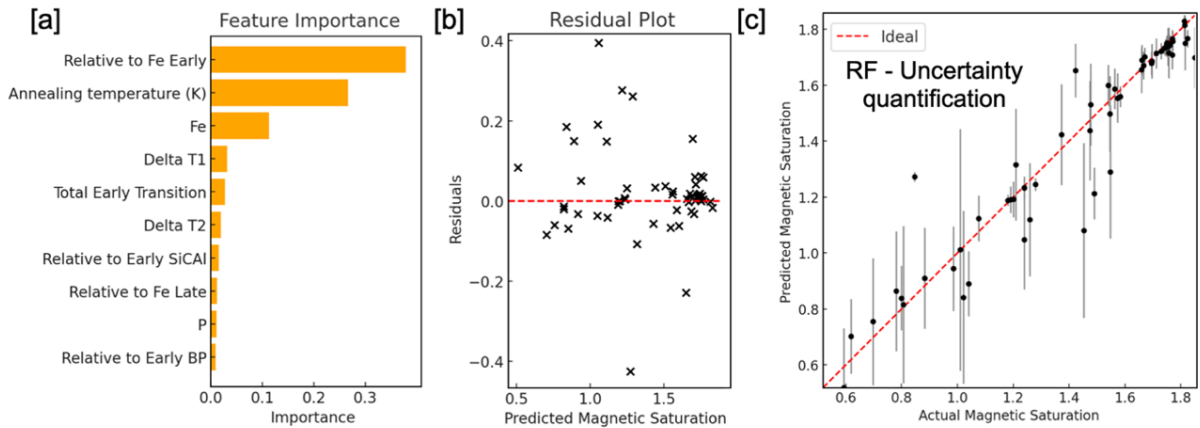

**Figure 3**. **(a)** Feature importance in predicting $M_S$ using a RF model. Features including *Relative to Fe Early*, *Annealing temperature (K)*, and *Fe* content exhibit the highest influence on the prediction. **(b)** The residual plot showing prediction error across the range of predicted magnetic saturation values. **(c)** Actual vs. predicted $M_S$ with associated uncertainty estimated from standard deviation across trees in the Random Forest ensemble done on original cross-validate data. Error bars reflect prediction uncertainty, with the red dashed line indicating the ideal one-to-one relationship.

Next, we evaluate the uncertainty in RF model in predicting $M_S$. **Figure 3c** represents the actual (true) values of magnetic saturation for each test sample, while the vertical axis represents the RF model's predictions for the same samples. The scatter plot compares actual versus predicted magnetic saturation values, where each data point is accompanied by an error bar representing the uncertainty of predictions across individual trees in the RF ensemble. RF model is an ensemble of N decision trees, and for a given feature vector x, each tree $T_i$ produces an estimate $y_i(x)$. The Random Forest prediction is taken to be the mean of these individual tree outputs, i.e., $\hat{y}(x) = \frac{1}{N} \sum_i^N y_i(x)$. To capture the RF model's uncertainty, we computed the variance of tree predictions: $\widehat{Var}(\hat{y}(x)) = \frac{1}{1-N} \sum_i^N [y_i(x) - \hat{y}(x)]^2$, while standard deviation of uncertainty in RF model was computed using formula $\sigma\,(\hat{y}(x)) =$ $\sqrt{Var(\hat{y}(x))}$. Each solid black circle in **Fig. 3c** denotes the mean predicted magnetic saturation for a particular Fe-alloy, and the vertical error bars around each dot depict the model's predictive uncertainty for that sample. Notably, the majority of predictions lie close to the red dashed line, which represents the ideal relationship (y=x), indicating a high degree of model accuracy. The R² score for the test set is approximately 0.86, the RMSE (RMSE) is 0.11, and the mean absolute error (MAE) is 0.064. The vertical spread of the error bars varies, with tighter uncertainties observed in regions with dense training data, while larger uncertainties appear in sparsely sampled regions or at the extremes of the composition-property space. This reflects the model's confidence and provides a quantitative measure of reliability for each prediction. Notably, for uncertainty quantification and parity plots (see SI **Fig. S4-S7**), we employed an 80/20 train-test split with random state (~42) to ensure reproducibility while full dataset is used for training the $M_S$ and $H_C$ models.

**Interpretation of RF model for magnetic saturation**: The partial dependence plots (PDPs) in **Fig. 4** reveal how the top eight most influential features affect magnetic saturation predictions in Fe-based alloys. *Relative to Fe Early* emerges as the strongest predictor, showing a clear positive nonlinear relationship that plateaus at higher values, indicating enhanced saturation with early transition elements up to an optimal limit. Annealing temperature (K) also shows a strong monotonic trend, suggesting that thermal processing, particularly in the 700-800 K range, significantly improves magnetic properties, likely through microstructural refinement.

Interestingly, the Fe content, though inherently magnetic, shows diminishing returns, highlighting that beyond a certain point, higher Fe does not proportionally enhance saturation, possibly due to compositional imbalance. Features like Delta T1 and Total Early Transition display non-monotonic trends, indicating optimal processing or alloying windows where magnetic performance peaks. In contrast, Relative to Early SiCAl and Relative to Fe BP exhibit weaker or plateauing effects, suggesting limited or even adverse impacts when present in excess. These plots collectively underscore the dominance of alloy chemistry and heat treatment in governing magnetic behavior and highlight the nonlinearity and thresholds critical for optimizing material design. (more data and details on uncetainity quantification and interpretation of RF model on test dataset used in next section is provided in supplemental figures **Fig. S4-S7**).

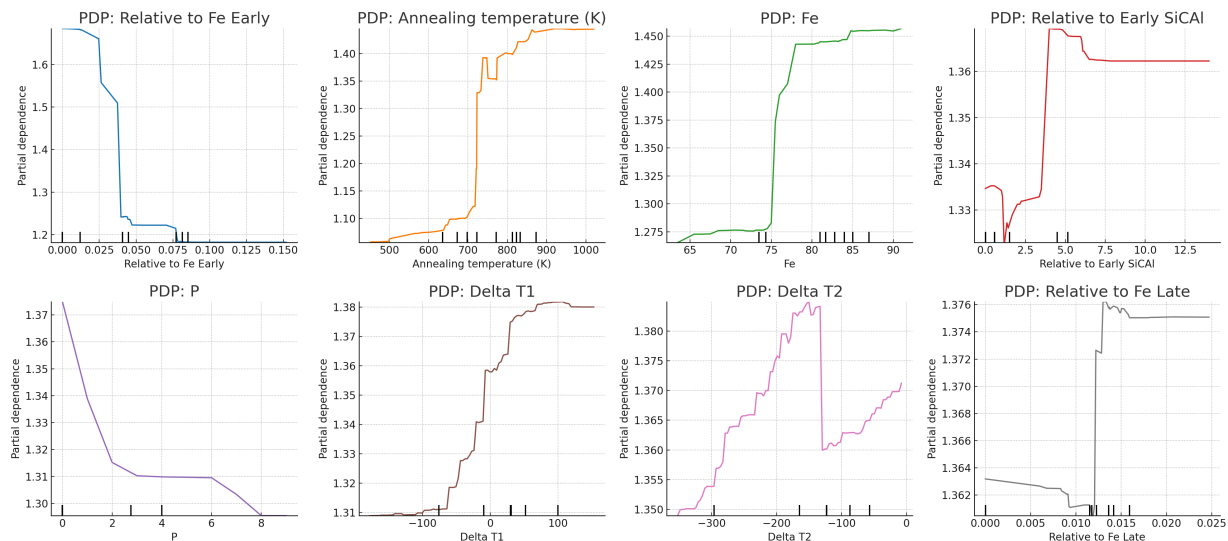

**Figure 4.** Partial Dependence Plots (PDPs) showing the marginal effects of the top eight most important features on predicted magnetic saturation in Fe-based alloys. The plots reveal key nonlinear relationships, with *Relative to Fe Early* and *Annealing temperature (K)* demonstrating the strongest positive influences. Other features, such as *Fe content*, *Delta T1*, and *Total Early Transition*, exhibit diminishing or optimal-response behaviors, highlighting the combined role of alloy composition and processing conditions in determining magnetic performance.

***ML model Prediction of magnetic saturation and validation from DFT and experiments***: We leveraged our best performing ML model (see **Fig. 2**), i.e., RF model, to perform a high-throughput analysis on ternary Fe-Si-B (**Fig. 5a,b**). To further validate the performance of our models, we evaluated their predictive alignment with experimental data using metrics such as the coefficient of determination ($R^2$), mean squared error (MSE), and mean absolute error (MAE). Both models—used for predicting magnetic saturation ($M_S$) and coercivity ($H_C$)—achieved high training $R^2$ values ($R^2 \approx 0.92$ for $M_S$ and $R^2 \approx 0.969$ for $H_C$) and low prediction errors, demonstrating their ability to capture known composition–property trends. The $M_S$ model built including elemental features like B, Cr, Cu, Fe, and Zr, showed strong agreement with experimental measurements and correctly captured the reduced in $M_S$ with increasing Si and B content. This makes good sense as Fe is a strong magnetic element, and reducing its content in the alloy should have direct impact. Notably, models that included unphysical or redundant features sometimes produced trends that disagreed with experiments, even when their overall accuracy looked good. This highlights that poor feature selection can lead to misleading predictions. A high $R^2$ does not always guarantee meaningful results if the features are not appropriate. In this context, $R^2$ is not just a performance metric, but a useful signal for judging whether a model has learned the right structure–property relationships. Choosing features with physical significance is essential when using ML to guide alloy design and validate results experimentally.

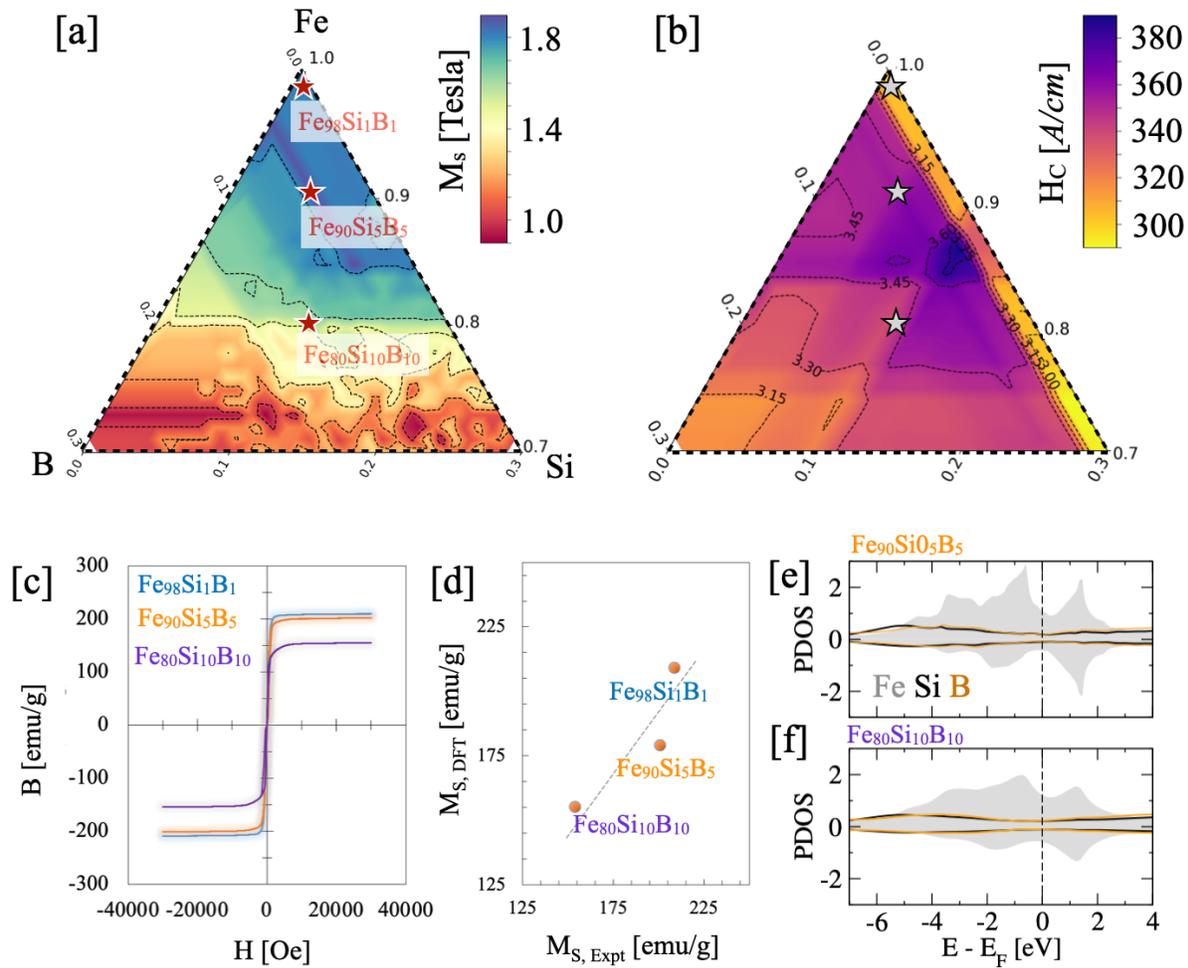

**Figure 5.** ML predicted ternary contour plots of (a) magnetic saturation ($M_S$; Tesla) and (b) coercivity ($H_C$; A/cm) for Fe-Si-B HEA. Both $M_S$ and $H_C$ show increase near Fe-rich regions while it decreases in Si and B regions. (c) Magnetic hysteresis loops for $Fe_{98}Si_1B_1$, $Fe_{90}Si_5B_5$ and $Fe_{80}Si_{10}B_{10}$ compositions, illustrating the effect of increasing Si and B on saturation magnetization. (d) Comparison between DFT-calculated and experimentally measured $M_S$, demonstrating a strong correlation between computational predictions and experiments. Electronic density of states of (e) $Fe_{0.90}Si_{0.05}B_{0.05}$, and (f) $Fe_{0.80}Si_{0.10}B_{0.10}$ showing reduced magnetic exchange with increasing Si-B concentration, in agreement with ML predictions and experiments.

The ideal alloy composition for high-performance soft-magnetic applications typically consists of 0.85-0.95 at.-frac. Fe, with Si and B concentrations in the range of 0.5-0.10 atomic percent each. Thus, the composition range of Fe and Si/(B-Cu/Nb/Zr) in **Fig. 5** is chosen as 07-1.0 and 0-0.3 at.-frac., respectively. The goal here is to identify composition-property relationship specifically targeting higher magnetic saturation and lower coercivity. The ternary contour plots in **Figure 5** illustrate the variation in magnetic properties, i.e., $M_S$ (**Fig. 5a**), and $H_C$ (**Fig. 5b**) as a function of ternary Fe-Si-B compositions. In **Fig. 5a**, a higher Fe content, particularly in the range of 85-100 atomic %, corresponds to increased saturation magnetization, as Fe is the primary magnetic element responsible for $M_S$. The introduction of Si and B generally reduces $M_S$, with Si contributing to the formation of $Fe_3Si$, which can lower the overall magnetic moment, while B enhances amorphization, further decreasing Ms. The highest $M_S$ values, around 1.6 Tesla, are observed in Fe-rich regions with minimal Si and B, while lower values appear in compositions with significant Si or B additions. Notably, as Si and B concentrations increase to a moderate range of approximately 10-20 atomic percent, lattice distortions are expected to emerge. These distortions weaken Fe-Fe exchange interactions, leading to a slight reduction in $M_S$. In addition to suppressing magnetization, excessive Si content also increases electrical resistivity, which is beneficial for reducing eddy current losses but further detracts from overall magnetization.

The $H_C$ plot in **Fig. 5b** reveals that the lowest coercivity values, around 300 A/cm, are found in Fe-rich regions, suggesting a soft magnetic behavior ideal for applications requiring low-energy losses. Notably, we found an increase in $H_C$ with Si and B content, which is attributed to expected change in microstructural features including grain refinement and increased defect density (SI **Fig. S8-S10**). The addition of B promotes the formation of an amorphous phase, which can lower coercivity in certain ranges but also introduce structural inhomogeneities that may lead to local increases in Hc. Similarly, Si can influence grain growth and domain wall movement, impacting coercivity. The highest coercivity values are observed in regions with balanced Si and B additions, where complex structural interactions likely impede magnetic domain motion. These observations highlight the critical role of composition-dependent microstructure in determining the magnetic performance of Fe-Si-B alloys.

The hysteresis curves in **Fig. 5c** for $Fe_{0.90}Si_{0.05}B_{0.05}$ and $Fe_{0.80}Si_{0.10}B_{0.10}$ compositions provide direct experimental evidence of how Si and B affect magnetization. Both samples exhibit a typical ferromagnetic response, with clear saturation behavior at high magnetic fields. $Fe_{0.90}Si_{0.05}B_{0.05}$ alloy exhibits a higher saturation magnetization (~200 emu/g), while $Fe_{0.80}Si_{0.10}B_{0.10}$ has a slightly reduced $M_S$ (~180 emu/g). The reduction in magnetization with increasing Si and B suggests that these elements either reduce the number of magnetic moments per unit volume or introduce structural modifications that lower overall magnetization. The coercivity values extracted from these loops further confirm the trend observed in the ternary diagram, where increased Si and B contribute to lower $H_C$, making the material magnetically softer.

In **Fig. 5d**, we compared experimentally measured saturation magnetization with DFT (see references [**117-125**]), which highlights the predictive accuracy of computational models. The data points closely follow a linear trend, indicating that DFT-based calculations reliably estimate experimental magnetization. While there are minor deviations, they likely stem from factors not accounted for in simulations, such as microstructural defects, phase impurities, or measurement uncertainties. The strong correlation suggests that DFT can be a useful tool for guiding alloy design before experimental synthesis, helping to predict the effects of compositional variations on magnetic properties.

We also analyzed the electronic-structure of $Fe_{0.80}Si_{0.10}B_{0.10}$ (**Fig. 5e**) and $Fe_{0.90}Si_{0.05}B_{0.05}$ (**Fig. 5f**), which reveals that iron dominates the electronic states near the Fermi level, with unfilled Fe-*3d* states playing a crucial role in magnetization through exchange interactions. While Si-*3p* shows minimal hybridization with Fe-*3d*, suggesting a limited direct impact on magnetic properties (it may suppress localized magnetic moments on Fe atoms, subtly influencing magnetostriction) while contributing to structural stability and reduced magnetostriction. The presence of Si and B, which disrupts the metallic bonding in Fe, could also hints towards increased electrical resistivity. B-*2p* exhibits slight hybridization with Fe-*3d*, modifying Fe *d*-orbital states and reducing magnetic anisotropy, which enhances soft magnetic behavior by lowering coercivity. The reduction in magnetic anisotropy due to B-2p and Fe-3d interactions not only lowers coercivity but also facilitates domain wall motion, further enhancing soft magnetic behavior [126]. As Si and B content increase in $Fe_{0.80}Si_{0.10}B_{0.10}$ (**Fig. 5f**), the overall PDOS near Fermi level decreases, indicating weaker magnetic interactions and a decline in magnetization due to electronic structure modifications. The observed smoothing of the PDOS at $E_F$ in $Fe_{0.80}Si_{0.10}B_{0.10}$ suggests increased electronic delocalization, further reducing Fe moment interactions and reinforcing the trend of decreasing magnetization. The reduced states available for magnetic exchange for $Fe_{0.80}Si_{0.10}B_{0.10}$ align with the observed decrease in saturation magnetization, demonstrating that while Si and B possibly refine grain size and improve soft magnetic properties, their higher concentrations lead to diminished magnetization. At higher Si and B concentrations, segregation to grain boundaries could also contribute to the reduction in magnetic interactions, reinforcing the observed decline in magnetization.

We also provide some down-selected soft-magnetic compositions for Fe-Si-(B-Cr/Cu/Nb/Zr) based materials class in **Table 1**. The table shows ML predictions and experimental validation for Fe-Si-B alloys, highlighting trends in magnetic and mechanical properties calculated using DFT. For $Fe_{80}Si_{10}B_{10}$, the ML-predicted $M_S$ of 1.57T closely matches the experimental value of 1.54T, demonstrating strong predictive accuracy. The $H_C$ is experimentally 289 A/cm, aligning with trends observed in ternary maps. Increasing Si and B content lowers $M_S$ and density, likely due to the dilution of Fe's magnetic moment. Moreover, the predictions for Fe-Si-B alloys with Cr, Cu, Zr, and Nb show minimal variations in $M_S$ (1.58–1.60 T), while coercivity increases slightly with Zr (320 A/cm) and Nb (330 A/cm), indicating higher magnetic hardness. Density remains relatively stable, with Zr and Nb-containing alloys reaching 7.11 g/cc. Notably, the bulk modulus increases significantly with Zr (2.41 Mbar), suggesting enhanced mechanical stiffness, making it a potential candidate for structural applications. The close match between ML and experimental values supports the use of computational methods for guiding alloy design before experimental validation.

**Table 1**. ML predicted Fe-rich compositions with optimal saturation magnetization and coercivity, which are compared with DFT calculated saturation (Tesla) [**115**], density (g/cc), and intrinsic-strength (GPa).

| System | $M_S$ [T] | | $H_C$ [A/cm] | Moment [T] | Density [g/cc] | Bulk Moduli [Mbar] |
|---|---|---|---|---|---|---|
| | ML | Expt. | ML | | | |
| **ML prediction, DFT and experimental validation** | | | | | | |
| **$Fe_{98}Si_1B_1$** | 1.81 | 2.09 | 301 | 2.04 | 7.78 | 188 |
| **$Fe_{90}Si_5B_5$** | 1.79 | 2.01 | 348 | 1.81 | 6.36 | 198 |
| **$Fe_{80}Si_{10}B_{10}$** | 1.57 | 1.54 | 333 | 1.54 | 5.34 | 122 |
| **Predictions** | | | | | | |
| **$Fe_{85}Si_{10}B_4Cu_1$** | 1.68 | | 290 | 1.65 | 5.91 | 186 |
| **$Fe_{84}Si_{10}B_4Zr_2$** | 1.60 | | 320 | 1.67 | 5.56 | 241 |
| **$Fe_{84}Si_{10}B_4Nb_2$** | 1.44 | | 330 | 1.63 | 5.92 | 198 |

Finally, we explored the four pseudo-ternary alloy combination, i.e., Fe-Si-(B-Cr) (**Fig. 6a,b**), Fe-Si-(B-Cu) (**Fig. 6c,d**), Fe-Si-(B-Nb) (**Fig. 6e,f**), and Fe-Si-(B-Zr) (**Fig. 6g,h**). **Figure 6** illustrates the variation in $M_S$ and $H_C$ for Fe-Si-B alloys upon the addition of Cr, Cu, Nb, and Zr. A quantitative analysis of the trends reveals significant differences in the extent of $M_S$ and $H_C$ reduction among these elements. In **Figure 6a** (Fe-Si-B-Cr), the addition of Cr causes a substantial decline in $M_S$, with values ranging from approximately 1.8T in Fe-rich regions to as low as 0.95T in Cr-rich regions. This decrease suggests that Cr disrupts ferromagnetic interactions, likely due to its lower magnetic moment and possible anti-ferromagnetic coupling with Fe. A similar but slightly less severe trend is observed in **Fig. 6c** (Fe-Si-B-Cu), where the saturation magnetization drops to a comparable minimum of around 0.94T. However, in Cu-containing alloys, Fe-rich regions retain relatively higher $M_S$, indicating that Cu's primary effect is dilution rather than strong magnetic interactions. On the other hand, in **Figure 6e** (Fe-Si-B-Nb), the impact on magnetization is even more pronounced, with M$_S$ values dropping from a maximum of 1.73T to a minimum of approximately 0.39T in Nb-rich compositions. This substantial reduction suggests that Nb strongly disrupts magnetic exchange interactions, possibly through increased structural disorder or electronic effects that weaken Fe-Fe coupling. Notably, Fe-Si-B-Zr system in **Fig. 6g** shows the least reduction in M$_S$, with values ranging from 1.85 T to 0.98T.

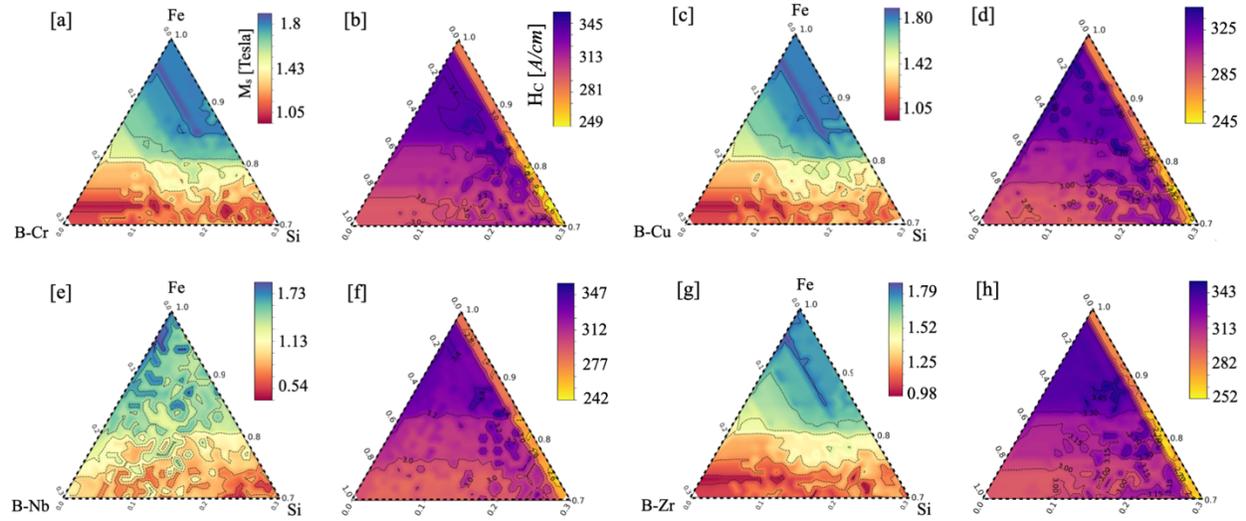

**Figure 6**. ML predicted magnetic saturation (in *Tesla*) and coercivity (in *A/cm*) of (a, b) Fe-Si-(B-Cr), (c, d) Fe-Si-(B-Cu), (e, f) Fe-Si-(B-Nb), and (g, h) Fe-Si-(B-Zr) over ternary/pseudo-ternary space showing possible trade-off between the two key properties. The range of Fe and Si/(B, Cr/Cu/Nb/Zr) composition ranges are chosen as 0.7-1.0 and 0-0.3 at.-frac., respectively.

Compared to Cr, Cu, and Nb, the addition of Zr appears to preserve a higher saturation magnetization, likely due to its minimal impact on Fe-Fe interactions. The relatively moderate decrease in M$_S$ suggests that Zr does not strongly interfere with magnetic ordering, making it a more favorable alloying addition for maintaining magnetic properties. Quantitatively, Cr and Cu lead to a decrease in M$_S$ to approximately 1.05T, while Nb has the most detrimental effect, lowering M$_S$ to 0.39T. Zr results in the least reduction, maintaining magnetization above 0.98 T. These findings indicate that alloying elements influence Fe-Si-B magnetization differently, with Cr and Nb causing the most significant reductions, Cu having a moderate impact, and Zr being the least disruptive to magnetic properties

While for H$_C$, the highest values are also observed near Fe-rich compositions, but the variation is less pronounced compared to magnetization. The Fe-Si-(B-Cr) (**Fig. 6b**) and Fe-Si-(B-Cu) (**Fig. 6d**) systems display relatively uniform coercivity distributions, with peak values near the Fe-rich corner. However, when Nb (**Fig. 6f**) and Zr (**Fig. 6h**) are introduced, the coercivity distribution becomes more irregular, suggesting that these elements contribute to changes in the microstructure that impact the magnetic domain behavior, which is also observed in feature analysis in SI **Fig. S1** (also see SI **Fig. S2** for coercivity) and SI **Fig. S9,S10**. Nb and Zr tend to slightly increase $H_C$ compared to Cu, which implies they may enhance grain refinement or introduce secondary phases that hinder domain wall motion, which is consistent with previous experimental observations.

Notably, Fe-Si-B-Cu offers the best soft magnetic performance with the highest Ms (1.85 T) and $H_C$ (290 A/cm) while Fe-Si-B is provided as a balanced baseline. Fe-Si-B-Zr and Fe-Si-B-Nb exhibit slightly lower $M_S$ but higher $H_C$, suggesting improved structural stability. Fe-Si-B-Cr shows slightly higher $H_C$ and lower $M_S$ (**Fig. 6a**), making it more suitable for corrosion-resistant applications rather than soft magnetics. We also compared our predictions with similar soft-magnetic literature on Fe-based nanocrystalline and amorphous alloys such as NANOMET ($Fe_{84.8}Si_{0.5}B_{9.4}Cu_{0.8}P_{3.5}C_1$) [127], FINEMET ($Fe_{73.5}Si_{13.5}B_9Cu_1Nb_3$) [128], NANOPERM ($Fe_{88}Zr_7B_4Cu_1$) [129], and HITPERM ($Fe_{44}Co_{44}Zr_7B_4Cu_1$) [130] have been developed, offering superior magnetic properties with $M_S$ values ranging from approximately 1.2 T to 1.94 T while low $H_C$. Similar to NANOMET, FINEMET, NANOPERM or HITPERM ML predicted compositions represent a balance between achieving high saturation magnetization and maintaining moderate coercivity, which is desirable for soft magnetic materials. The presence of Nb and Zr slightly increases coercivity compared to Cu, but they might also enhance structural stability, making them suitable for applications where a trade-off between $M_S$ and $H_C$ is required.

### 4. Conclusion

In this study, we developed an interpretable, data-driven machine learning framework to predict magnetic saturation ($M_S$) and coercivity ($H_C$) in Fe-based soft magnetic alloys, with the goal of accelerating the discovery and optimization of Co- and Ni-free materials. The Random Forest (RF) model, trained entirely on curated experimental datasets, achieved high predictive accuracy and exhibited strong agreement with in-house measurements across the Fe-Si-B compositional space. Notably, the model captured the observed decrease in $M_S$ from ~2.01 T to ~1.54 T as Si and B concentrations increased, a trend validated by both experimental measurements and density functional theory (DFT) calculations. The DFT analysis confirmed that higher Si and B content introduces significant disorder broadening in the electronic density of states (DOS), reducing magnetic moment density and weakening Fe-Fe exchange interactions.

Beyond accurate prediction, the RF model facilitated high-throughput screening of extended compositional families, including Fe-Si-X (X = B, Nb, Zr, Cu) and pseudo-ternary systems such as Fe-Si-(B-Cu), Fe-Si-(B-Nb), and Fe-Si-(B-Zr). The model identified promising alloy compositions with optimized trade-offs between high $M_S$ and low $H_C$, informed by interpretable feature importance and partial dependence analysis. Key processing parameters such as annealing temperature, Delta T1/T2, and crystallization dynamics were found to significantly affect magnetic performance. This integrative ML–DFT approach not only improves prediction fidelity but also offers physical insights into the role of composition and thermal history, establishing a robust approach for the rational design of next-generation soft magnetic materials.

## 5.  Acknowledgement

A.N. is grateful for the research opportunity at Ames National Laboratory supported by the Iowa State University-Ames National Laboratory's Science Research program. The Ames National Laboratory is supported by the U.S. Department of Energy (DOE), Office of Science, Basic Energy Sciences, Materials Science and Engineering Division. The research was performed at the Ames National Laboratory, which is operated for the U.S. DOE by the Iowa State University under contract No. DE- AC02-07CH11358.